# Direct Mapping of Intrinsic Topology of Bound States in the Continuum via Nonlinear Emission

*Shuzheng Chen[1,#], Hongwei Wang[2,#], Zijian He[1,#], Liyu Zhang[1], Kai Wang[1,3,*], Xu Jiang[2], Jiaxing Yang[1], Yuda Wan[1], Guangwei Hu[2,*] and Peixiang Lu[1,4,*]*

[1]Wuhan National Laboratory for Optoelectronics and School of Physics, Huazhong University of Science and Technology, Wuhan 430074, China

[2]School of Electrical & Electronic Engineering, Nanyang Technological University, 639798, Singapore

[3]School of Electronic and Information Engineering, Hubei University of Science and Technology, Xianning 437100, China

[4]Hubei Key Laboratory of Optical Information and Pattern Recognition, Wuhan Institute of Technology, Wuhan 430205, China

#These authors contributed equally to this work.

*Corresponding authors:

kale_wong@hust.edu.cn, guangwei.hu@ntu.edu.sg, lupeixiang@hust.edu.cn

**Abstract:**

The direct mapping of the intrinsic topology in a leaky photonic band is crucial and challenging in topological photonics. For instance, observables in bound states in the continuum (BICs) feature complex topological textures such as a polarization vortex in momentum space, which nonetheless is difficult to be characterized in far-field scattering, especially considering the dominant direct channel. Here, we propose and experimentally demonstrate a hybrid nonlinear metasurface that enables a direct visualization of the intrinsic topology in BICs via second-harmonic generation (SHG). The enhanced local-source of SHG from the ultrathin indium tin oxide can effectively excite the emissions from the eigenmodes of a $TiO_2$ photonics crystal slab, achieving three-order enhancement of SHG magnitudes. Importantly, these enhanced SH emissions carry topological polarization textures of BICs to the far field. With this, we can directly construct polarization vector maps of symmetry-protected BICs and chiral symmetry-broken quasi-BICs, clearly visualizing the winding structure around V points, the generation and evolution of chiral C points. This work provides a universal approach for characterizing topological photonic systems via coherent nonlinearity processes, opening new avenues for studying topological phenomena in non-Hermitian photonic systems.

## Introduction:

Bound states in the continuum (BICs) are the counterintuitive optical modes within leaky photonic systems, arising from the destructive interference between different radiative components. It facilitates diverging quality factors (Q factors) in theory and topologically protected polarization singularities (V points) in $k$-space[1–4], which enables various applications such as low-threshold lasers[5–8], high-sensitivity sensors[9], vortex beam generators[10], and full-polarization controllers[11]. Active manipulation of BICs has been obtained via merging multiple BICs in $k$-space, drastically boosting their quality factors and robustness[12–16]. Alternatively, symmetry breaking splits charges to form circularly polarized states (C points)[17,18]. These principles underpin recent advances like intrinsic chiral BICs[19,20], maximally chiral emission[21–23], Janus BICs[24], and efficient chiral metasurfaces[25], greatly expanding capabilities for light-matter control based on BICs.

Nonetheless, the direct probing of the intrinsic topology of BICs in $k$-space remains a critical and ongoing challenge. The conventional far-field scattering technique faces inherent limitations in resolving complex polarization topologies with repeated measurements, thus to construct polarization vector maps in $k$-space[3,26,27]. This inevitably requires precise polarization control and measurement, extensive data sampling and post-processing. Alternatively, the emissions of BICs can transform the topological polarization textures to far fields, suggesting the direct probe of intrinsic topology through a single-shot measurement of, for example, fluorescence[28,29], lasing[6,7,30], and thermal emission[31] from BIC modes. However, due to the incoherent nature, both fluorescence and thermal emission are not suitable for topological polarization mapping. While BICs lasing shows a high coherence, it always requires sufficient high-Q optical modes to satisfy pump thresholds, which may only work for the BICs with high-symmetry. In addition, the external gain media could remarkably influence the optical properties of BICs, increasing the difficulties in design and fabrication.

Second-harmonic generation (SHG) offers a direct path to overcome these limitations[32,33]. As a coherent emission process, it preserves the full polarization information necessary for topological mapping. Furthermore, its threshold-free operation enables direct probing of eigenmodes across momentum space without relying on high-Q lasing or external gain media. Reports of enhanced SHG from BICs underscore this potential[34–37]. To implement such a scheme, a highly efficient local-source is critical. This is provided by an indium tin oxide (ITO) layer, which leverages its epsilon-near-zero (ENZ) region to significantly boost SHG efficiency[38–40], thus serving as an ideal coherent local-source.

In this work, we present a direct mapping of intrinsic topology in BICs through second-harmonic (SH) emissions, in a platform of ultrathin ITO layer loaded by a $TiO_2$ photonic crystals slab (PCS), as shown in Fig. 1a. Herein, SHG can be enhanced jointly by its ENZ properties and BICs modes in $TiO_2$ PCS, exhibiting a three-order of magnitudes of SH enhancement. More importantly, as a coherent nonlinear process, SH emission can carry the intrinsic topology of BICs, thus allowing the direct mapping of polarization singularities of BICs in $k$-space. We hence showcase the reconstruction of polarization vector maps of symmetry-protected BICs and symmetry-broken q-BICs, thus clearly resolving the winding structure around V point and the evolution of C point in $k$-space. Our work establishes the bridge between the intrinsic topology of BICs and far-field polarization vector maps, providing a universal approach for characterizing topological photonic systems.

## Principle of hybrid nonlinear metasurface for BICs Topology

The intrinsic topology of BICs is encoded in SH emissions, forming a specific far-field polarization vector map of SHG in *k*-space. Obviously, it is crucial to enhance SHG conversion efficiency for this localized nonlinear probing technique. Therefore, the hybrid nonlinear metasurface is designed to support double resonances[41,42] (Fig. 1b). Specifically, the wavelength of the pump light is tuned to be matched with the ENZ region[38] of ITO layer, and then the out-of-plane electric component $z^E_\omega$ within the ITO layer is significantly amplified and results in the SHG enhancement. Accordingly, the resonance of BICs in TiO$_2$ PCS is designed to be at half of ENZ wavelength, and then the SH local-source is effectively coupled to the eigenmodes. Therefore, the SH emission is further enhanced and carries the topological polarization information to the far-field. The total SH conversion efficiency can be expressed as[41,43]:

$$W_{2\omega,\vec{k}} \propto Q_{\vec{k}} L_{\vec{k}}(2\omega) \kappa_{12,\vec{k}} \left[ \kappa_1 \left( W_{\omega,\vec{k}'} \right) \right]^2 \tag{1}$$

$W_{\omega,\vec{k}'}$ is the input power in Fourier space. $\kappa_1(W_{\omega,\vec{k}'})$ describes the nonlinear polarization with in-plane vector $\vec{k}$ benefits from field enhancement and large second-order susceptibility of thin layer ITO. Here, $\kappa_{12,\vec{k}}$ is the cross-coupling coefficient which depends on the spatial overlap integral between nonlinear polarization and the eigenmode with in-plane wave vector $\vec{k}$. The spectral overlap factor is defined as $L_{n\vec{k}}(2\omega) = \frac{\gamma_{n\vec{k}}^2}{\left(\omega - \omega_{n\vec{k}}\right)^2 + \gamma_{n\vec{k}}^2}$. This indicates that emission enhancement does not occur when the frequency is in off-resonance condition. $Q_{n\vec{k}}$ is the quality factor of the eigenmode, and a higher Q factor leads to a greater SH emission enhancement. A coupled mode theory for scattering and emission is appended in Section 1 and 2 in Supplementary Information, SI for detailed discussions.

For the demonstration, Fig. 1c shows normalized SH intensities (red triangles) from the bare ITO ultrathin layer (10 nm) by tuning the excitation wavelength (1200 nm~1300 nm). Clearly, due to ENZ effect, SH enhancement by one-order magnitudes is observed around 1230 nm (blue area), in comparison to that in the off-resonance region, which agrees with recent literature reports. Furthermore, atop the ITO layer, we fabricate a TiO$_2$ PCS with symmetry-protected BICs at 615 nm, i.e. SH wavelength (see Method and SI for sample fabrication details). Surprisingly, over 500 times enhancement of SH intensity from the TiO$_2$/ITO nonlinear metasurface (blue pentagrams) is achieved in comparison to that from the bare ITO layer (resonant pumping at 1240 nm). Note that the pumping wavelength is tuned to 1240 nm for TiO$_2$/ITO nonlinear metasurface, which is slightly off-resonance from the Γ-point of BICs. As a result, the total SH emission is significantly enhanced over three order-of-magnitude compared to that in off-resonance region, providing a stable and convenient light source for the topological probe (see Method and SI Section 4-6 for characteristics and enhancement of SHG). More importantly, this strong SH emission carries topological polarization information, establishing the bridge between the intrinsic band topology of BICs and far-field polarization vector maps.

To clarify fundamental differences between SH emission and linear scattering, Fig. 1d shows a home-built Fourier-based spectroscopy system. It allows switching between emission and scattering modes while keeping all other optical and polarization settings identical. By introducing another SH reference beam, interference patterns can be obtained for analyzing topological charge. In the left panel of Fig. 1e, a clear single-armed spiral pattern is observed, indicating the vortex beam with a topological charge of $l = 1$. Nonetheless, for the linear scattering mode, a different double-armed spiral pattern is observed as shown in the right panel of Fig. 1e, indicating a topological charge of $l = -2$. This exactly reveals the profound differences in mechanism between these two modes: for linear scattering mode, BICs act as a phase plate in $k$-space that convert the incident light into a vortex beam, whereas SH emission corresponds to the inherent topology of BICs, enabling the direct mapping of its intrinsic band topology. (See Section 1 and 2 in SI for details)

## Intrinsic Topology in Symmetry-Protected BICs

We further investigate topological properties of symmetry-protected BICs in TiO$_2$/ITO nonlinear metasurface by using Fourier-based nonlinear imaging system (see Method and SI for details). Firstly, the excitation wavelength is tuned to 1240 nm to match the ENZ region of the ITO layer, and the SH wavelength thus coincides with the eigenmodes around the BIC in the TiO$_2$ PCS. Fig. 2a presents the measured image of the SH intensity distribution ($S_0$) in $k$-space, indicating a clear doughnut-shaped isofrequency contours. And then, the polarization states of the SH emission are fully characterized, and the measured images of Stokes parameters, $S_1, S_2$ and $S_3$ are presented in Fig. 2b. Note that all these parameters are normalized by $S_0$. Based on these Stokes parameters, the BICs polarization vector maps in $k$-space can be fully constructed. The polarization ellipse for each pixel is defined by its orientation angle ($\psi$) and ellipticity angle ($\chi$), which can be extracted from obtained Stokes parameters by $\psi = \frac{1}{2}\arctan\left(\frac{S_2}{S_1}\right)$ and $\chi = \frac{1}{2}\arcsin\left(\frac{S_3}{S_0}\right)$. The chirality of the ellipse is determined by the sign of $S_3$ ($S_3 > 0$ for LCP, $S_3 < 0$ for RCP). The region with a high degree of linear polarization (DOLP⩾0.85) are highlighted in black.

Fig. 2c presents the resulted SH polarization vector map, revealing a perfect polarization vortex structure centered at $\Gamma$ point, which shows a $2\pi$ polarization angle variance and thus confirms a topological charge of $q = 1$. Under off-resonance excitation conditions at 1300 nm, the SH intensity distribution exhibits notable differences compared to the near-resonance case as shown in Fig. 2d. Although the spatial distributions of the normalized Stokes parameters, $S_1/S_0$ and $S_2/S_0$, are different, they exhibit a two-fold rotational symmetry, and the values of $S_3/S_0$ remain close to zero, indicating the dominance of linear polarization throughout $k$-space as shown in Fig. 2e. In Fig. 2f, it indicates that despite under the off-resonance condition, it still maintains a winding polarization angle distribution consistent with that under the near-resonance condition

In-depth analysis reveals that experimental results under near- and off-resonance conditions essentially correspond to measurements of isofrequency contours of the same photonic band at different frequencies. The differences of distribution patterns in *k*-space are entirely determined by the intrinsic dispersion characteristics of photonic band, while the SHG intensity variations essentially stem from the changes in enhancement effects: under the near-resonance condition, the synergistic enhancement effect significantly boosts the nonlinear conversion efficiency; under off-resonance conditions, both enhancement effects are diminished, leading to the attenuation of the SH intensity. Nevertheless, the SH intensity is still very stable for measurement, enabling the reconstruction of the polarization vector map under the off-resonance conditions. (see Section 7 in SI for details) These results fully demonstrate unique advantages of the proposed nonlinear detection method: it is not only able to accurately characterize singularities by leveraging the synergistic enhancement effect under near-resonance but also can acquire generalized topological textures across the entire *k*-space by varying the excitation wavelength. This capability to resolve the complete topological structure of the *k*-space provides a solid foundation for all relevant research efforts in this field.

**Intrinsic Topology in Symmetry-Broken BICs**

Our nonlinear probing technique proves equally capable of resolving the complex polarization maps of quasi-BICs due to symmetry breaking. To actively manipulate the topological charge of BICs, the specific $TiO_2$ PCS with a structural modification is fabricated. Fig. 3a shows the SEM image of the PCS composed of modified meta-atom. The modification involved the selective removal of the top-right quadrant, which reduced the structural symmetry from $C_{4v}$ to $C_{2v}$. In Fig. 3b, the angle-resolved transmission spectrum reveals that the resonant wavelength of the BICs is slightly shifted, and its Q-factor is decreased in comparison with that of the symmetry-protected sample in Fig. 2 due to the symmetry-broken effect. Fig. 3c presents the measured *k*-space images of SH intensity distribution ($S_0$) of the symmetry-broken sample under the near-resonance pumping at 1240 nm. Unlike the doughnut-shaped distribution of the symmetry-protected BICs in Fig. 2a, no obvious intensity singularity is observed near the $\Gamma$ point, indicating that the original V point has vanished. Furthermore, the extracted intensity distributions of the LCP and RCP components (Fig. 3d) are complementary in *k*-space, indicating two separate extremum points. They do not coincide with the $\Gamma$ point but exhibit a symmetric distribution about the $\Gamma$ point, heralding the generation of a pair of C points.

Fig. 3e presents the obtained normalized Stokes parameter, $S_1/S_0$, $S_2/S_0$ and $S_3/S_0$. Specifically, the distribution of $S_2/S_0$ parameter exhibits a two-fold symmetry characteristic related to the system's $C_{2v}$ symmetry. A distinction with positive-value regions along the specific direction in *k*-space is formed, reflecting the intrinsic symmetry axes after symmetry breaking. Importantly, the parameter of $S_3/S_0$ shows a significant distribution of non-zero value: predominantly negative in the first quadrant (corresponding to RCP) and predominantly positive in the third quadrant (corresponding to LCP). Clearly, this spatial separation and sign variation serve as the key evidence for the generation of C points. Fig. 3g shows the reconstructed polarization map, which is in good agreement with the theoretical simulations in Fig. 3f. Thus, it confirms that the V point at the $\Gamma$ point has vanished, and a pair of C points have emerged. C points possess an identical half-integer topological charge but opposite chirality, appearing symmetrically in the first and the third quadrants. This result demonstrates that symmetry breaking enables the splitting of topological charges from integer to half-integer.

## Splitting and Evolution of Integer Topological Charges in BICs

Furthermore, we systematically study the evolution of chiral C points across varying degrees of symmetry breaking. Fig. 4a presents the design of unit cell: a cylindrical nanopillars with a diameter of 248 nm, featuring a sector-shaped notch at its top-right corner, where the size of the notch is defined by the parameter $\Delta$, representing the radius of the sector. This value can be continuously adjusted thus enabling full control of the degree of symmetry breaking. Fig.4b shows SEM images of samples with three different $\Delta$ values: 60 nm, 80 nm, and 100 nm. Clearly, as the $\Delta$ value is increased, the structural asymmetry gradually enhances. Furthermore, the measurements of angle-resolved transmission spectra indicate that as $\Delta$ is increased, the Q-factor of the BICs decreases, confirming the enhancement of the symmetry-breaking effect (see Fig. S5 in SI for details).

In terms of nonlinear optical characterization, we study the dynamic evolution of topological states under different degrees of symmetry breaking via simulations (Fig. 4c). It shows that, as the asymmetry parameter $\Delta$ increases, a pair of C points with opposite chirality has emerged from the $\Gamma$ point and gradually evolve along symmetric directions. This can be explicitly mapped and reconstructed via SH emission (Fig. 4d), which are in good agreement with simulations. Along the symmetry axis connecting a pair of C points, two distinct flips in the major-axis direction are observed[17]. The locations of these flip points directly identify specific coordinates of C points in k-space, providing a robust experimental criterion for their localization. When $\Delta$ = 60 nm (with a local enlarged view inserted left), the C points (mark with red and blue dots respectively) are close to the $\Gamma$ point; as $\Delta$ is increased to 80 nm and 100 nm, the C point pairs show significant separation, and their movement trend is consistent with the simulation results.

Our experimental results intuitively show the splitting and evolution of topological charges, offering clear evidence for actively manipulating topological photonic states via geometric design. This finding fills the gap in the quantitative correlation between symmetry breaking degree and topological singularity dynamic evolution, advances topological photonics' control from qualitative realization to precise control, and provides a key experimental paradigm for subsequent in-depth studies on the intrinsic coupling mechanisms and dynamic behaviors of topological states.

## Conclusion：

In summary, we report a novel local nonlinear probing paradigm for the intrinsic topology of BICs using a TiO$_2$/ITO nonlinear metasurface. It supports over three-order of magnitude of SHG enhancement based on a double-resonance mechanism. Importantly, the polarization characteristics of the enhanced SHG enable direct visualization of the intrinsic topology of BICs. Using this platform, we have not only accurately identified the winding structure of V points in symmetry-protected BICs but also, for the first time, systematically tracked the evolution of chiral C points in symmetry-broken quasi-BICs under controlled symmetry breaking conditions. Furthermore, it shows the ability for the detection of other eigenmodes.

Therefore, it shows the potential applications in non-Hermitian topological photonic systems[44–46] such as topological skyrmions[47] and exceptional points[26,48–51]. This study pioneers a new pathway for experimental research on topological light fields and lays an important foundation for the development of nonlinear topological photonics.

## Methods:

**Sample fabrication**: A commercially available $SiO_2$ substrate pre-coated with a 10-nm-thick ITO layer was used. A 140-nm-thick $TiO_2$ film was subsequently deposited via magnetron sputtering. The $TiO_2$ nanopillar array was patterned using electron-beam lithography (EBL, Raith ELINE Plus) on a layer of PMMA 950K resist. After development, a Cr hard mask was deposited by electron-beam evaporation and structured via lift-off in acetone. The pattern was transferred into the $TiO_2$ layer by ICP etching under an $SF_6$ atmosphere at 8 mTorr and 80 W RF power. Finally, the remaining Cr mask was removed using a commercial Cr etchant solution. (See Fig. S6 in SI for details)

**Optical measurement:** As illustrated in Fig. S7 in SI, nonlinear optical characterization was performed using a tunable femtosecond laser system (Insight DS DUAL, 680-1300 nm, 120 fs, 80 MHz) as the excitation source. The incident polarization was controlled using an infrared QWP (WPQW-IR-4M, Sigma Koki; 1000-1600 nm). A custom-built Fourier optics-based system was employed for signal generation and detection. Fundamental wave focusing was achieved through a near-infrared objective lens (Mitutoyo, 50×, 0.42 NA), creating a diffraction-limited focal spot. Emitted nonlinear signals were collected by an objective lens (Olympus, 20×, 0.40 NA) and directed through a calibrated 4f optical relay system to a scientific CMOS camera (Prime 95B, Photometrics). Spectral filtering was implemented using a 720 nm short-pass filter combined with a 500 nm long-pass filter to isolate the SHG signal. For polarization-resolved measurements, a visible-range QWP (WPQW-VIS-4M, OptoSigma) and a linear polarizer (FLP25-VIS-M, LBTEK) were inserted before the detector to analyze the polarization state.

**Numerical simulation:** Numerical simulations are performed in COMSOL Multiphysics eigen-frequency solver to compute the band structures and eigenpolarizations. Periodic boundary conditions are applied in the *x* and *y*-directions. To simulate eigenmodes, perfectly matched layers (PMLs) are implemented along the *z*-direction to absorb outgoing radiation. The distribution of nonlinear polarization of ITO in both of real space and *k*-space are calculated using MATLAB. (See Section 1-5 in SI for details)

## Acknowledgements

This work was supported by National Key Research and Development Program of China (No. 2022YFA1604403) and National Natural Science Foundation of China (No. 12274157, No. 12021004, No. 12274334, No. 11904271), Natural Science Foundation of Hubei Province of China (No. 2023AFA076). Special thanks are given to the Analytical and Testing Center of HUST, the Experiment Center for Advanced Manufacturing and Technology in the School of Mechanical Science & Engineering of HUST, and the Center of Micro−Fabrication and Characterization (CMFC) of WNLO for use of their facilities. G. H. acknowledges the start-up grant from Nanyang Technological University, Ministry of Education (Singapore) under AcRF TIER1 (RG61/23) and AcRF TIER2 (MOE- T2EP50125-0038), National Research Foundation of Singapore under award no. NRF-CRP31-0001, and A*STAR under its MTC YIRG Grant (Project No. M23M7c0119) and MTC IRG Grant (Project No. M24N7c0081, Project No. M24N7c0087).

## Data availability

The main data supporting the findings of this study are available within the article and its Supplementary Information files. Extra data are available from the corresponding author upon reasonable request.

## Author contributions:

K. W., G. H., P. X. L. conceived the project. K.W., G. H., P. X. L. supervised the project. S. Z. C., H. W. W. and Z. J. H. designed the experiments. S. Z. C., Z. J. H and Y. D. W performed the experiments. L.Y.Z. performed theoretical calculations and numerical simulations. S. Z. C., H. W. W., Z. J. H, L.Y.Z., J. X. Y., X. J., Y. D. W analyzed data. All authors discussed the results. S. Z. C., L.Y.Z. and K. W. drafted the paper with the inputs from all authors.

## Competing interests:

The authors declare no conflicts of interest.

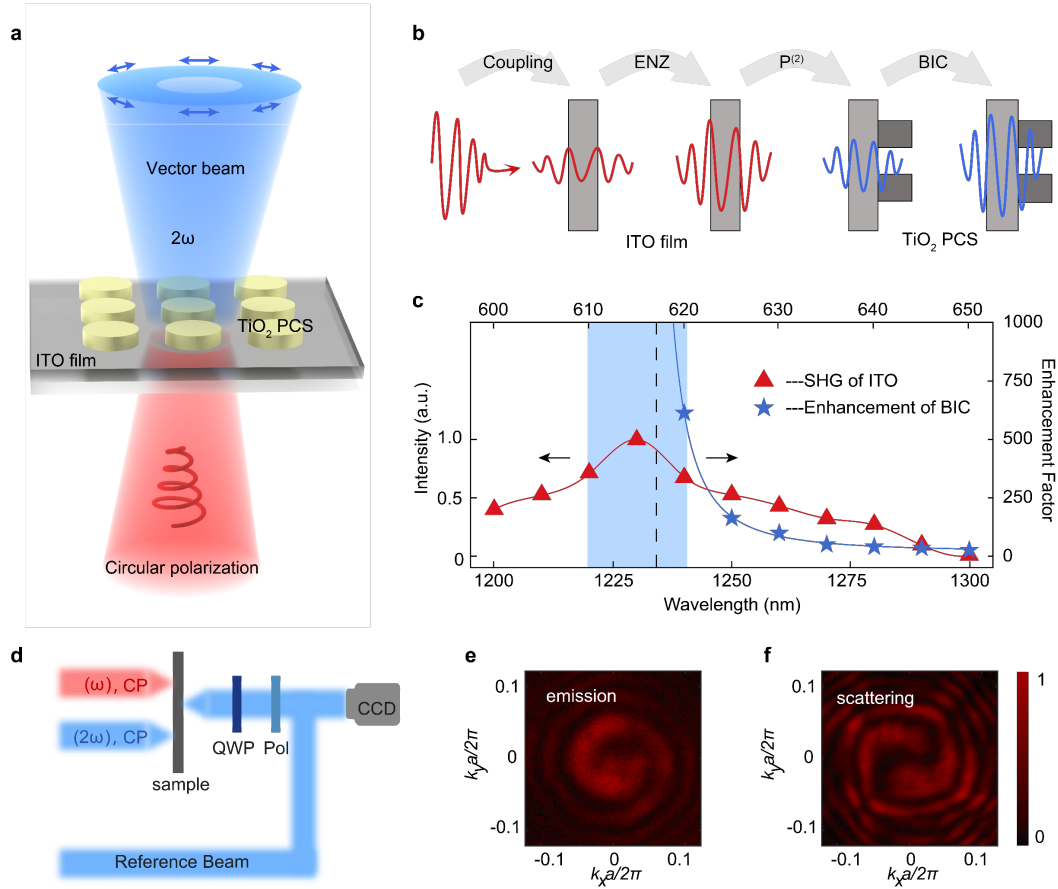

**Figure 1 | Principle of probing intrinsic topological charges via locally excited SH emissions.** (a) The hybrid metasurface, comprising a TiO$_2$ PCS on an ultrathin ITO film (10 nm), converts a circularly polarized fundamental light into SH vector beams carrying intrinsic topology in *k*-space. The PCS consists of cylindrical nanopillars with a diameter of 248 nm, a height of 140 nm, and a lattice period of 400 nm. (b) Schematic of the SHG process in the metasurface: The fundamental field is enhanced in the ITO layer at its ENZ region, generating a strong nonlinear source. This source then couples to and excites eigenmodes in the TiO$_2$ PCS at the SH wavelength. (c) Verification of double-resonance-enhanced SHG. The normalized SH intensity from the bare ITO layer (red triangles) shows an order-of-magnitude enhancement around 1230 nm due to the ENZ effect; the hybrid metasurface (blue pentagrams) exhibits a remarkable enhancement of over 500 times relative to the bare ITO film when pumped at 1240 nm. (d) The experimental set-up: allows switching between emission and scattering modes while keeping all other optical and polarization settings identical. (e) The interference pattern of the emission shows a single-arm spiral structure, confirming a topological charge of $l = 1$. (f) The interference pattern of the scattering shows a double-arm spiral structure, confirming a topological charge of $l = -2$.

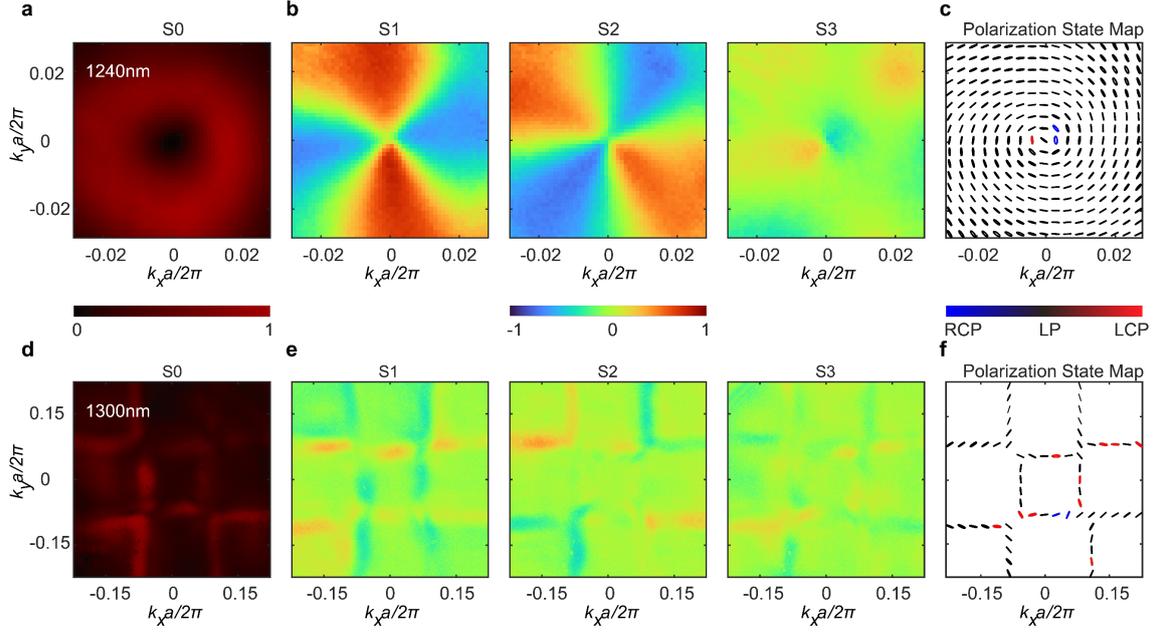

**Figure 2 | Topological texture of a symmetry-protected BIC revealed via nonlinear SH mapping.** (a–c) SHG response under near-resonance excitation (at 1240 nm), where the synergistic enhancement between the high Q-factor of the BIC and the extreme field enhancement in the ITO's ENZ region provides intense nonlinear emission. (a) SHG intensity ($S_0$) distribution showing a canonical doughnut-shaped profile. (b) Normalized Stokes parameters ($S_1/S_0$, $S_2/S_0$ and $S_3/S_0$) indicating a purely linear polarization state ($S_3/S_0 \approx 0$) across $k$-space. (c) Reconstructed polarization map revealing a vortex with a continuous +1 winding of the polarization vector around the $\Gamma$ point, confirming an integer topological charge (V point). Black regions denote a high degree of linear polarization (DOLP ≥ 0.85). (d–f) SHG response under off-resonance excitation (at 1300 nm). Despite a significantly weaker intensity, the +1 topological winding structure around the Γ point is preserved, demonstrating the method's ability to resolve the intrinsic photonic band topology beyond the resonance condition.

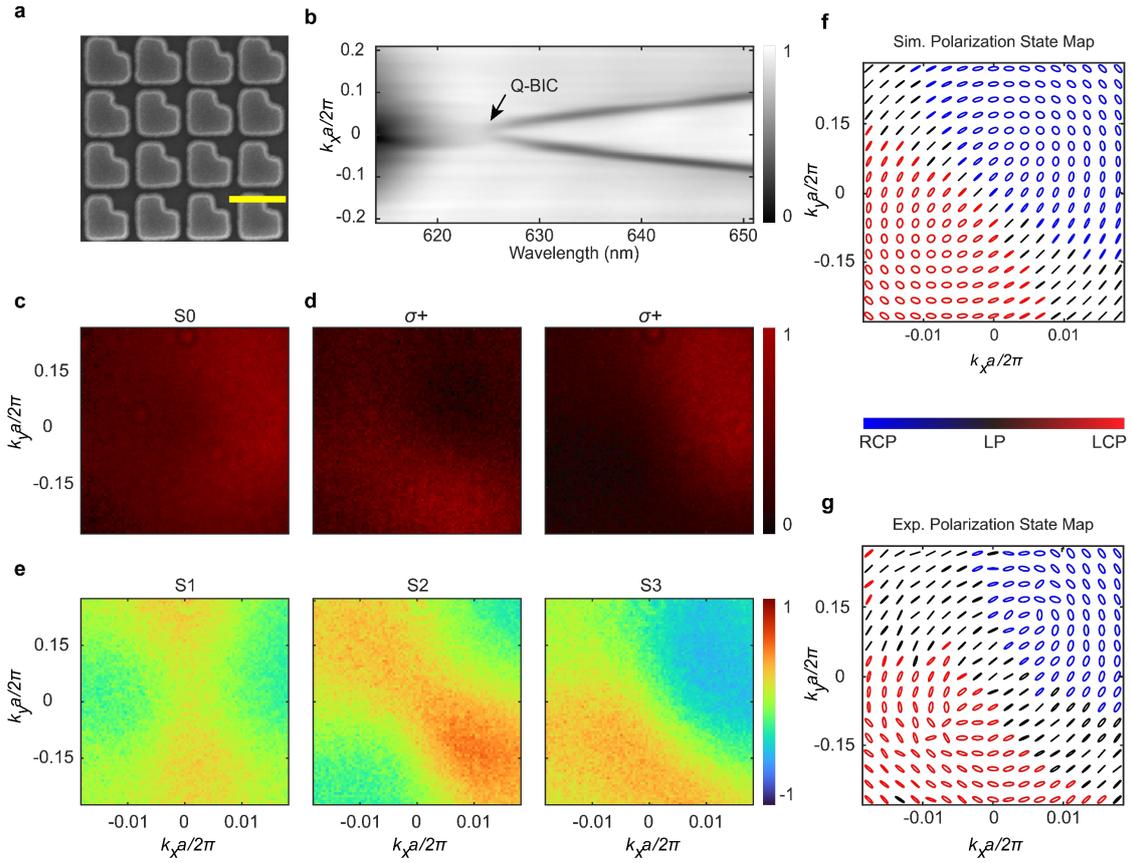

**Figure 3 | Topological texture of a symmetry-broken qBIC revealed via nonlinear SHG mapping.** (a) Top-view scanning electron microscopy (SEM) image of the fabricated symmetry-broken hybrid metasurface. Scale bar, 500 nm. (b) Angle-resolved transmission spectrum of the sample, with the black arrow indicating the q-BIC resonance. (c) SHG intensity distribution ($S_0$) in $k$-space under near-resonance pumping. The disappearance of the intensity singularity at the $\varGamma$ point indicates the annihilation of the original V point. (d) Circular polarization-resolved SHG intensity distributions. The spatially separated intensity maxima for left-handed (LCP, left) and right-handed (RCP, right) components are symmetric about $\varGamma$, heralding the formation of a pair of C points. (e) Normalized Stokes parameters ($S_1/S_0$, $S_2/S_0$ and $S_3/S_0$). (f, g) Excellent agreement between the (f) theoretically calculated and (g) experimentally reconstructed polarization maps, conclusively demonstrating the topological evolution: the BIC is destroyed, and the integer topological charge splits into a pair of half-integer charges, corresponding to C points with opposite chirality.

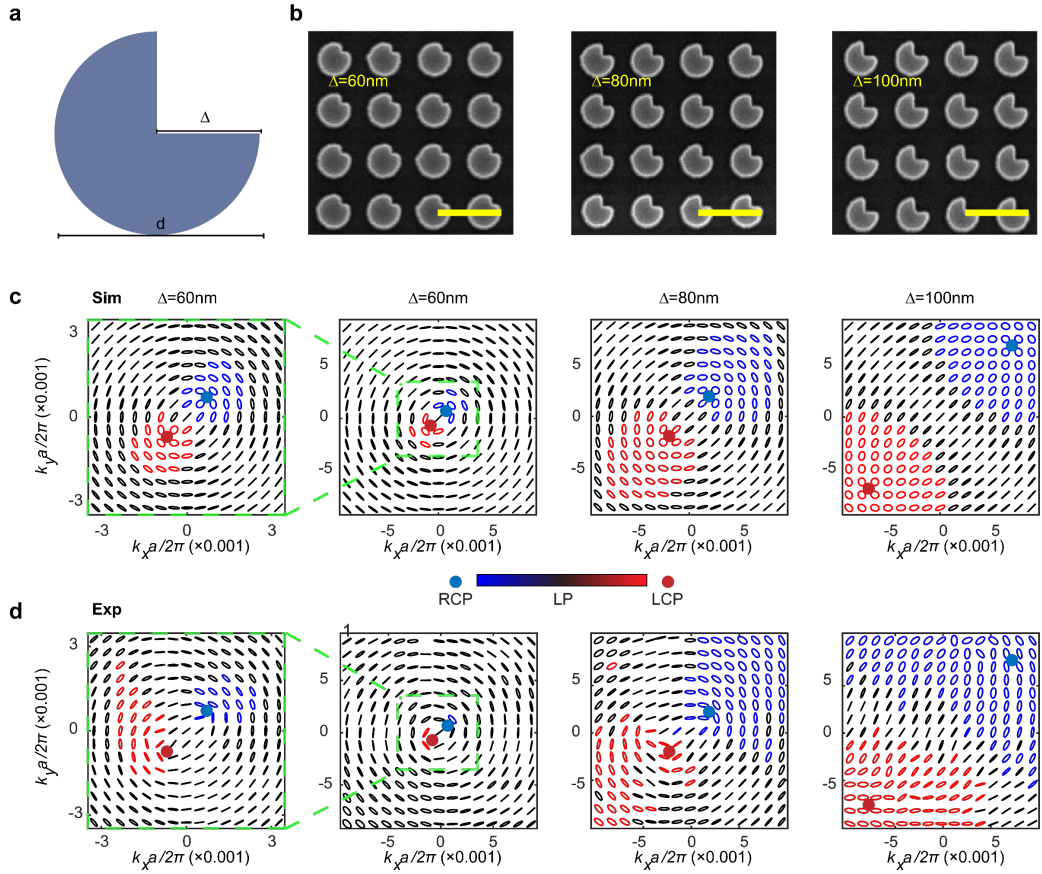

**Figure 4 | Tracking the evolution of C points with opposite chirality via nonlinear SHG mapping.** (a, b) Unit cell design with a tunable sector notch (Δ= 60 nm, 80 nm and 100 nm) and corresponding SEM images, demonstrating precise control over the degree of asymmetry. Scale bars: 500 nm. (c, d) Evolution of the polarization maps as a function of the sector radius parameter Δ from (c) simulation and (d) experiment. As Δ increases, a pair of C points with opposite chirality emerge near the $\Gamma$ point and symmetrically separate toward the periphery, unambiguously revealing the dynamic evolution of topological polarization singularities.